# Observation of "volume reflection" effect in crystal collimation experiments

V.M. Biryukov[♦]

*Institute for High Energy Physics, Protvino, Russia*

**Abstract**

Strong effect of beam coherent scattering ("reflection") in a field of bent crystal is observed in crystal collimation experiments performed with heavy ions and protons at RHIC and started at Tevatron collider. Detailed simulation using Monte Carlo code CATCH is done in order to understand the observations and relate them to the physics of beam coherent scattering in crystal. A.M. Taratin and S.A. Vorobiev predicted the effect of beam «volume reflection» in bent crystals in 1987. The presented data is the first manifestation of this new physical phenomenon in experiment.

Bent-crystal technique [1] is well established for extracting high energy beams from accelerators. It was successfully applied at up to world highest energy [2], and simulations were able to predict the results correctly [3]. Experiments at IHEP Protvino [4] have demonstrated that this technique can be quite efficient: 85% of the beam have been extracted using a short (2 mm) Si channeling crystal with bending of about 1 mrad, with intensity of the extracted 70-GeV beam up to $3\times10^{12}$ protons per spill. Much of the IHEP physics program relies on crystal channeled beams used regularly since 1989 [5].

It would be promising to apply the bent-crystal technique for a beam halo scraping at high-energy colliders [6,7]. A bent crystal, serving as a primary element, should coherently bend halo particles onto a secondary collimator. A demonstration experiment of this kind was performed in 1999 at IHEP where a factor of 2 reduction in the accelerator background was obtained with a bent crystal incorporated into beam cleaning system [8].

The theory of crystal extraction and collimation is based mainly on detailed Monte Carlo simulations tracking the particles through a curved crystal lattice and the accelerator environment in a multipass mode [9,10]. Simulation code CATCH [11] was successfully tested in extraction experiments at CERN SPS [12], FNAL Tevatron [13], BNL RHIC [14], and IHEP U-70 [4]. Monte Carlo predictions, suggesting a "multipass" mode of crystal extraction where efficiency is dominated by the multiplicity of particle encounters with a short crystal, were a clue to the breakthrough in the extraction efficiency demonstrated at IHEP Protvino [5].

A possibility to improve beam halo scraping using a bent channeling crystal instead of a thin scattering primary collimator has been studied in detailed realistic simulations for the Tevatron [6] and Large Hadron Collider [15]. It was shown that the scraping efficiency can be substantially increased and the accelerator-related backgrounds in the Tevatron and LHC collider detectors can be reduced by about one order of magnitude.

Another experiment on crystal collimation (of Au ions and protons) has been in progress at the Relativistic Heavy Ion Collider in collaboration of BNL with IHEP [16]. The yellow ring of the RHIC had a bent crystal collimator, 5 mm along the beam. By properly aligning the crystal to the beam halo, particles entering the crystal were deflected away from the beam and intercepted downstream in a copper scraper. RHIC crystal collimator efficiency measured for gold ions as a

---
[♦] http://mail.ihep.ru/~biryukov/

function of the crystal angle was found in good agreement with CATCH simulations with the measured machine optics, as seen in Fig. 1. For the 2003 RHIC run, the theory predicted the efficiency of 32%, and averaging over the data for this run gave the measured efficiency of 26% [17]. The modest figure of efficiency ≈30%, both in theory and experiment, is attributed to the high angular spread of the beam that hits the crystal face as set by machine optics. It is worth to compare this figure of efficiency for gold ions at RHIC to the 40% efficiency achieved with the same (5 mm O-shaped) crystal for protons at IHEP in 1998 [18]. It is also worth to notice that the crystal extraction efficiency observed at CERN SPS with Pb ions was 4-11% with a long (40 mm) crystal of silicon [19]. Following the RHIC experiment, the first crystal collimator is now installed into Tevatron, making use of the same crystal taken from the RHIC crystal collimation set-up [20].

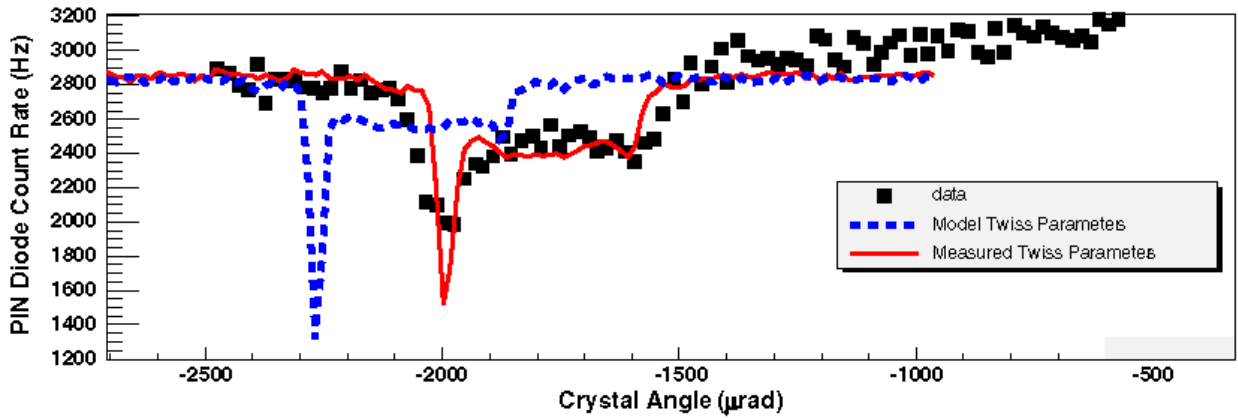

**Figure 1** The rate of nuclear interactions in the crystal measured (dots) and simulated as a function of the crystal orientation at RHIC [14]. Two sets of CATCH simulation shown: with preliminary optics (dash blue) and with measured optics (solid red).

In understanding the crystal collimation experiments, major attention was paid to the peak efficiency of crystal channeling which makes applications so attractive. However, apart from the strong channeling peak, collimation experiment has also shown an interesting (and unexpected originally) distinct feature - a very substantial reduction in the background rate observed over a broad angular range corresponding to the bending angle of the crystal (0.44 mrad), between -2000 and -1500 μrad in Fig. 1. Interestingly, the numerical experiment, i.e. CATCH *ab initio* simulations, nicely reproduced this "shoulder" in the collimation plots [14]. The simulations assumed a bent crystal lattice of atoms with Moliere model applied to the potential of individual atoms; particles were tracked through the lattice with tiny steps; see [1] for more details. While the agreement is good, the interpretation of the results remained an important issue. This strong manifestation of some unexpected coherent crystal effects, observed in a broad angular range, is interesting by itself, but also can lead to serious consequences in crystal applications at accelerators.

In order to understand this phenomenon, we did further detailed simulations for RHIC and Tevatron experiments, studying the particle dynamics in single and multiple interactions with bent crystal in different angular ranges in the environment of collimation experiment. For a crystal at random orientation to the beam, these interactions are found to be very similar to those in amorphous media. For a certain range of crystal orientation, a particle can become tangential to atomic planes somewhere in the bent crystal depth. Variety of effects in these conditions is known from the physics of channeling and quasi-channeling: "volume capture" (scattering induced transfers of random particles to channeled states) and "volume reflection" (scattering of random particles off the potential of bent atomic planes).

The volume capture phenomenon is well known from experiments [21] and theory [1]. Its probability at high energy is pretty small; we find it about 0.09% per one encounter with the crystal in case of RHIC experiment. The volume reflection phenomenon was predicted by A.M.

Taratin and S.A. Vorobiev in 1987 [22]; however, it was never observed experimentally so far. Fig. 2 (a) shows schematically the idea of particle reflection off the coherent field of bent atomic planes in crystal.

The simulated exit angular distributions for a particle (100 GeV/u gold ion for RHIC and 980 GeV proton for Tevatron) interacting with the crystal used in the collimation experiments are shown in Fig. 2 (b). These distributions show a clear shift in the average exit angle to the side opposite to crystal bending - this is the effect known as volume reflection. The average angular shift found in simulations is –15.6 µrad for RHIC and –4.2 µrad for Tevatron, i.e. on the order of 1-1.5 critical channeling angles. This shift is independent of the incidence angle within the range of crystal bending, between -2000 and -1500 µrad in Fig. 1. Apart from a spectacular shift, the distributions show a substantial broadening. The r.m.s. exit angle in Fig.2(b) is 18.2 (RHIC) and 6.4 (Tevatron) µrad. These angles should be compared to the r.m.s. angle of multiple scattering of the same particles at a random incidence at the crystal, i.e. outside of the conditions for channeling or reflection, namely to 12.9 µrad for RHIC and 3.3 µrad for Tevatron. Thus, a coherent scattering on the potential of bent atomic planes appears a very significant factor for particle dynamics in accelerator.

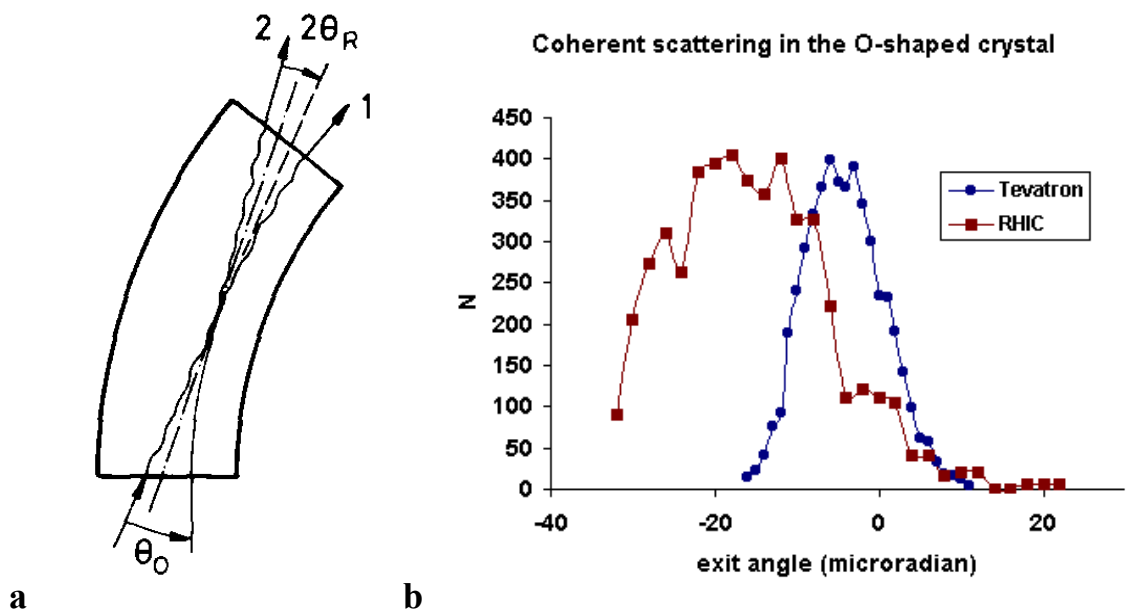

**Figure 2 (a)** Schematic picture of a channeled particle 1 and reflected particle 2 [22] **(b)** Simulated exit angular distributions for a particle (100 GeV/u Au ion for RHIC and 980 GeV proton for Tevatron) interacting with the crystal used in the collimation experiments. Zero angle corresponds to the direction of incident particle.

In a circular accelerator, the scattered particles continue the circulation in the ring and encounter the crystal again and again on later turns. The angular (and spatial) distribution of the particles changes due to betatron oscillations in the ring and due to scattering on every encounter with the crystal. For crystal orientation of -2000 to -1500 µrad in Fig. 1, the conditions for coherent scattering ("volume reflection") take place on every encounter with the crystal. For random orientation, this never happens. Respectively, the particle amplitude in accelerator grows faster if coherent scattering contributes strongly to overall scattering. By every encounter with a crystal, beam emittance grows by about $\beta(\Delta\theta)^2$, where $\beta$ is accelerator beta function and $\Delta\theta$ is the scattering angle in crystal. We have found that in RHIC case the emittance grows faster by factor of 2 in the plateau region of Fig. 1 than outside of this region.

This difference in beam dynamics on the phase space with crystal under conditions of coherent scattering leads to faster particle loss on the secondary elements in the accelerator. As a result, this effect reduces the particle loss (nuclear interactions) at the crystal itself, as part of the loss now goes to different elements in the ring. Therefore, if one observes the background rates downstream of a bent crystal in accelerator, they should reduce sizably when coherent scattering takes place in beam interaction with crystal. This provides explanation to the plateau effect in collimation plots like Fig.1 - a strong contribution from coherent fields to overall scattering strongly affects the pattern of particle loss along the accelerator and therefore strongly affects the local loss rate downstream of the crystal. It is not so important that coherent scattering produces literally "reflections" (i.e. the average angle deviates from 0); but the resulting increase in the overall r.m.s. angle of scattering in crystal from coherent effects is important, so the beam diffusion increases several-fold. This diffusion and the presence of the secondary aperture make the effect of plateau.

In order to check our understanding and make further quantitative predictions, we simulated in detail the Tevatron collimation experiment. More details of the settings used can be found in ref. [6]. We used the recent edition of Tevatron accelerator lattice [23]. In the plane of bending, the accelerator functions were $\beta$=60.205 m and $\alpha$=-0.216 at the crystal location. The crystal was placed at $5\sigma$ and served as a primary element in collimation scheme. The secondary collimator was placed 31.5 m downstream, at $5.5\sigma$. Particle tracking in Tevatron lattice is done with linear transfer matrices. Each particle was allowed to make an unlimited number of turns in the ring and of encounters with the crystal until a particle either undergoes a nuclear interaction in the crystal or hits the secondary collimator (either because of bending effect in the channeling crystal or because of scattering events). Overall approach to channeling simulation in the ring was like in [24]. A non-channeling amorphous layer 2 micron thick was assumed on the crystal surface due to its irregularity at a micron level.

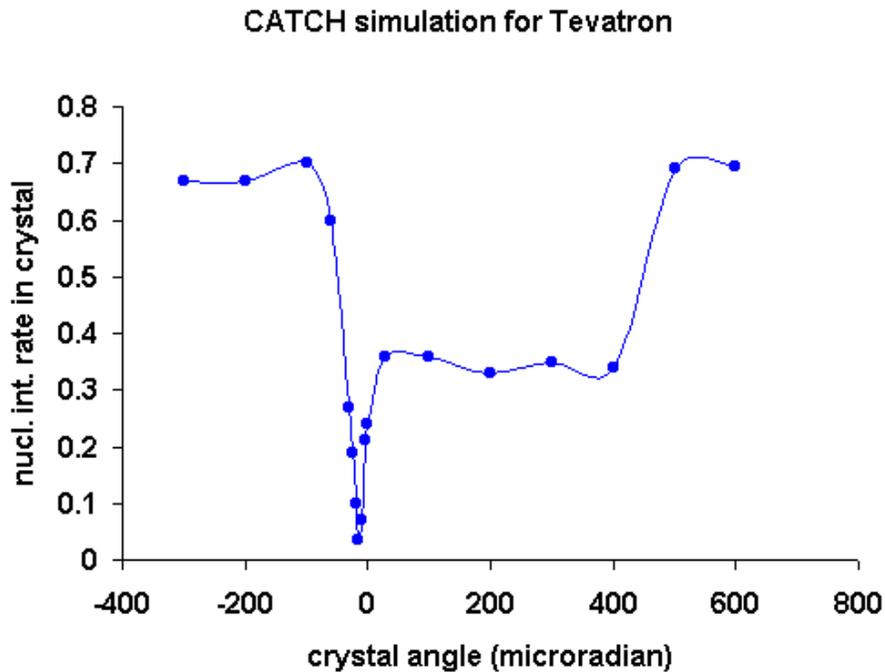

**Figure 3** The rate of nuclear interactions in the crystal predicted as a function of the crystal orientation at Tevatron.

The nuclear interaction rate in the crystal is shown in Fig. 3 as a function of crystal orientation at Tevatron. This rate was measured in RHIC experiment as shown in Fig. 1, and now this data is being taken in the ongoing collimation experiment at Tevatron. The remarkable dip on the plot, about -95% down from the rate observed at random orientation, is due to channeling with high efficiency predicted in the environment of Tevatron collimation experiment. One can expect about an order of magnitude reduction of machine-related backgrounds in the collider detector if crystal components are installed in the collimation systems of collider. The width of this dip found in simulations is about 30 μrad (while crystal critical angle is ±5 μrad), therefore the efficiency depends strongly on the crystal alignment. The channeling peak width is greater than critical angle because here channeling is essentially multipass, multiturn effect. Protons encounter crystal several times, scatter, go on circulating in the ring, and then get channeled on some later encounter. This scattering contributes sizably to the width of the peak.

Another distinct feature is the expected plateau on the plot, i.e. a strong reduction (by 50%) in the rate within the angle range of crystal bending, 0.44 mrad wide. We find that, within this range of crystal alignment, beam emittance grows by a factor of 3.7 faster than at random alignment, in Tevatron case. Our simulations show that at the 7-TeV LHC this factor can be as high as 10-20.

We can conclude that crystal collimation studies, besides promising a very high efficiency of the technique in colliders, reveal a new interesting physics of beam reflection on the coherent field of the atomic planes of bent crystal. This reflection, theoretically predicted almost 20 years ago, causes a strong perturbation of beam in the conditions of crystal collimation experiments at RHIC and Tevatron and is observed as a very strong factor affecting particle loss in the accelerator ring. This revealed physics is essential in designing crystal applications for colliders. This strong scattering effect could even serve itself as a basis of a new collimation design.

The author thanks N.V. Mokhov, D. Still, Y.M. Ivanov, and R. Fliller for useful discussions.

# References

[1] V.M. Biryukov, Yu.A. Chesnokov and V.I. Kotov, *"Crystal Channeling and its Application at High Energy Accelerators".* Berlin: Springer (1997)
[2] C.T. Murphy et al., Nucl. Instrum. Meth. B **119**, 231 (1996).
[3] V. Biryukov. Phys. Rev. E **52**, 6818 (1995).
[4] A.G. Afonin et al. Phys. Rev. Lett. **87**, 094802 (2001).
[5] A.G. Afonin, V.T. Baranov, V.M. Biryukov, V.N. Chepegin, Yu.A. Chesnokov, Yu.S. Fedotov, A.A. Kardash, V.I. Kotov, V.A. Maisheev, V.I.Terekhov, E.F. Troyanov. Nucl. Instr. and Meth. B **234**, 14 (2005)
[6] V.M. Biryukov, A.I. Drozhdin, N.V. Mokhov. PAC Proc. (New York, 1999) p.1234
[7] V.M. Biryukov [ArXiv:physics/0307027]
[8] A.G. Afonin et al. PAC Proc. (New York, 1999) p. 53
[9] V. Biryukov, Nucl. Instrum. and Meth. B **53**, 202 (1991)
[10] A. Taratin et al., Nucl. Instrum. and Meth. B **58**, 103 (1991).
[11] V. Biryukov. CERN SL/Note 93-74 AP (1993). "CATCH 1.4 User's Guide".
[12] H. Akbari et al. Phys. Lett. B **313,** 491 (1993)
[13] R.A. Carrigan, Jr., et al. Phys. Rev. ST AB **1**, 022801 (1998).
[14] R.P. Fliller, A. Drees, D. Gassner, L. Hammons, G. McIntyre, S. Peggs, D. Trbojevic, V. Biryukov, Y. Chesnokov, V. Terekhov, Nucl. Instr. and Meth. B **234**, 47 (2005).
[15] V.M. Biryukov, V.N. Chepegin, Yu.A. Chesnokov, V. Guidi, W. Scandale, Nucl. Instr. and Meth. B **234**, 23 (2005).
[16] R.P. Fliller, A. Drees, D. Gassner, L. Hammons, G. McIntyre, S. Peggs, D. Trbojevic, V. Biryukov, Y. Chesnokov, V. Terekhov. Phys. Rev. ST AB **9**, 013501 (2006)
[17] R.P. Fliller, A. Drees, D. Gassner, L. Hammons, G. McIntyre, S. Peggs, D. Trbojevic, V. Biryukov, Y. Chesnokov, V. Terekhov, AIP Conf. Proc. **693**, 192 (2003).
[18] A.G. Afonin et al., Phys. Lett. B **435**, 240 (1998).
[19] G. Arduini et al. Phys. Rev. Lett. **79,** 4182 (1997)
[20] N. Mokhov and D. Still, CERN Crystal Collimation meeting (Dec. 2005).
[21] V.A. Andreev et al., JETP Lett. **36**, 415 (1082)
[22] A.M. Taratin and S.A. Vorobiev. Phys.Lett. A **119** (1987) 425
[23] D. Still, private communication.
[24] V. Biryukov. Phys. Rev. Lett. **74**, 2471 (1995).